\documentclass[12pt,a4paper,dvips]{article}
\usepackage{a4p}
\usepackage{cite,mcite}
\usepackage{graphicx}
\usepackage{physics}
\usepackage{l3_title,ifthen}
\usepackage{Lep}
\usepackage{amssymb,amsmath}
\usepackage{epsfig}
\journalname{Phys. Lett. B}
\date{January 24, 2000}
\preprint{2000-018}
\newlength{\capindent}
\newlength{\capwidth}
\newlength{\figwidth}
\newcommand{\icaption}[2][!*!,!]{\hspace*{\capindent}%
  \begin{minipage}{\capwidth}
    \ifthenelse{\equal{#1}{!*!,!}}%
      {\caption{#2}}%
      {\caption[#1]{#2}}
  \end{minipage}}
\newcommand{\nngg}{\ensuremath{\nu\bar\nu\gamma(\gamma)}} 
\newcommand{\snu}{\ensuremath{\tilde{\nu}}}
\newcommand{\charg}{\ensuremath{\tilde{\chi}^{\pm}_1}}
\newcommand{\neut}{\ensuremath{\tilde{\chi}^0_1}}
\newcommand{\chichi}{\ensuremath{\tilde{\chi}^+_1 \tilde{\chi}^-_1 }}

\newcommand{\dm}{\ensuremath{\Delta \mathrm{M} \ }}
\newcommand{\dmns}{\ensuremath{\Delta \mathrm{M}}}
\newcommand{\m}{\mathrm{M}}
\newcommand{\tanb}{\ensuremath{\tan \beta}}

\newcommand{\dg}{\ensuremath{^{\circ}}}
\newcommand{\pba}{pb$^{-1}$ }

\newcommand{\sqs}{\ensuremath{\sqrt{s}} }

\begin{document}
\bibliographystyle{l3style}
\begin{titlepage}

\title{Search for Charginos with a Small Mass Difference to 
the Lightest Supersymmetric Particle at {\boldmath \sqs= 189 \GeV{}} }
\author{The L3 Collaboration}

%
%
\begin{abstract}
A search for charginos nearly mass-degenerate with the lightest
supersymmetric particle is
performed using the 176 \pba of data collected at 189 \GeV{} in 1998 with the L3
detector. Mass differences between the chargino and the lightest supersymmetric particle
below 4 \GeV{} are considered.
The presence of a high transverse momentum 
photon is required to single out the signal from the photon-photon
 interaction background. No 
evidence for charginos is found and upper limits on the cross section for 
chargino pair production are set. For the first time, in the case of
heavy scalar leptons, chargino mass limits are obtained for any 
$\charg - \neut$ mass difference.
\end{abstract}
%
%
\submitted

\end{titlepage}

%
%

\section{Introduction}
 Supersymmetry (SUSY) is one of the most appealing extensions of 
the Standard Model.
In SUSY theories with minimal particle
content (MSSM) \cite{mssm}, in addition to the ordinary particles, there is
a supersymmetric spectrum of particles with spins which differ by one
half with respect to their Standard Model partners.

Charginos (\charg{}), which are a mixture of the supersymmetric
 partners of \Wpm\, (gaugino state) 
and H$^{\pm}$ (higgsino state), are pair produced via $s$-channel 
$\gamma/\mathrm{Z}$ exchange. The production cross section can be
reduced by an order of magnitude when the $t$-channel scalar neutrino
(\snu{}) exchange is important.

In this paper the hypothesis of R-parity conservation is made.
The R-parity is a quantum number which 
distinguishes ordinary particles from supersymmetric particles.
If R-parity is conserved supersymmetric particles are 
pair-produced and their decay chain always contains,
besides standard particles, two Lightest Supersymmetric Particles (LSP).
The LSP, which is stable and weakly-interacting, escapes detection.
The LSP can be the lightest neutralino (\neut{}), 
a mixture of the supersymmetric partners of Z, \gam, and neutral Higgs bosons, 
or the scalar neutrino.

As long as charginos are sufficiently heavier than the LSP, their
decay products can be detected with high trigger and selection efficiencies.
When the mass difference ($\dmns = \m_{\charg} - \m_{\neut}$)
is smaller than 4 \GeV, the search described in 
Ref. \cite{stsearch} (hereafter referred as standard) becomes very 
inefficient, because in that range the signal and the photon-photon interaction
background are indistinguishable.

For $\dm$ between a few hundred \MeV{} and a few \GeV, charginos
decay near the interaction vertex and the energy carried by the 
visible decay products is so small that trigger inefficiencies become
substantial. In addition, such a signal is overwhelmed by the photon-photon
interaction background, which is rapidly increasing for decreasing masses
of the photon-photon system. The trigger efficiency and
the signal to background ratio can be improved by requiring in the
event an  Initial State Radiation (ISR) photon with high transverse
momentum. Therefore, in this paper we report a search for the process
$$ \ee \ \to \ \gamma \chichi \ \to \ \gamma  \neut 
  \neut \mathrm{X} $$ 
where X stands for low energy standard charged particles. 

This method was previously used in the Mark II experiment
 \cite{mark2} in the search for a fourth lepton doublet, whose
members are close in mass. It was then suggested \cite{gsearch} 
for the search at LEP for charginos nearly mass-degenerate with 
the LSP. The DELPHI collaboration recently published such a search in a 
data sample collected at
$\sqs\leq 183 \GeV$ \cite{DELPHI183}.

\section{Data Sample and Monte Carlo Simulations}
\label{simu}

This search uses the data collected at 189 \GeV{} with
the L3 detector\cite{L3-DETECTOR} for an integrated luminosity of 176 \pba.
The following background processes are simulated: $ \ee \to  
\tau^+ \tau^- \gamma(\gamma)$, $ \ee \to  \mu^+ \mu^- \gamma(\gamma)$ and 
$ \ee \to  \nngg $
with {\tt KORALZ}\cite{KORALZ}, $ \ee \to  {\rm W^+  W^-}\gamma (\gamma)$ 
with {\tt KORALW}\cite{KORALW} and radiative Bhabha scattering 
with {\tt TEEGG}\cite{TEEGG}.
The statistics of the Monte Carlo is equivalent
to more than 60 times the integrated luminosity except for radiative
Bhabha events where it is more than 16 times.

Photon-photon interactions with hadronic final state ($\ee \to \ee {\rm q
  \bar{q}}$) can not be reliably simulated
in the range of interest of this analysis, because of the
large theoretical uncertainties on the differential cross section 
in the non-perturbative regime.
For photon-photon interactions with leptonic final states 
($\ee \to \ee \ell^+\ell^- \gamma(\gamma)$) there is no complete Monte Carlo 
including the simulation of initial state radiation.
For this reason, no photon-photon background is used.

Signal events are simulated with {\tt SUSYGEN} \cite{susygen2.2} for
chargino masses between 45 \GeV{} and 88 \GeV{} and \dmns{} from 30 \MeV{} to 4 \GeV.
Only events with a photon more than 10\dg{} away from the beam pipe and 
with an energy greater than 4 \GeV{} are considered. These requirements
will be referred as fiducial cuts. 
The chargino decay
branching ratios as in Ref. \cite{gsearch} are used. 
For \dmns{} smaller than 200 \MeV, \charg{} decay
lengths of a few centimetres may occur. In this case, \charg{} decays are
treated by the {\tt GEANT} \cite{geant} package. 

The detector response is simulated using the {\tt GEANT} 
package. It takes into account effects of energy loss,
multiple scattering and showering in the detector materials and
in the beam pipe. Hadronic interactions are simulated with the
\texttt{GHEISHA} program \cite{gheisha}. Time dependent inefficiencies
of the different subdetectors and of the trigger are also taken into account
in the simulation procedure.

\section{Low {\boldmath \dmns} phenomenology}\label{pheno}

The search for low \dmns{} charginos accompanied by ISR photons
does not suffer from the large photon-photon interaction background if
the transverse momentum of the photon is large enough.
While in signal events the missing momentum is due to weakly interacting
particles, in photon-photon interactions it is due to the electrons escaping
in the beam pipe. In the latter case, if a high transverse momentum photon is
present the two final state electrons must be deflected into the detector.
To suppress the photon-photon interactions, the following requirement on the
photon transverse momentum is applied:
\begin{equation}
 E_{T \gamma} \geq 2 E_{\rm beam} \frac{\sin \theta_d}{1+\sin \theta_d} \label{etcut}
\end{equation}
where $\theta_d$ is the minimum detection angle for the deflected electron 
and $E_{\rm beam}$ is the beam energy. For the L3 detector, 
$\theta_d = 1.7\dg$
resulting in $E_{T \gamma} \geq 5.45 \GeV$. With this requirement 
the photon energy is large enough to trigger the detector.

\begin{figure}[p]
\begin{center}
\includegraphics[height=15cm]{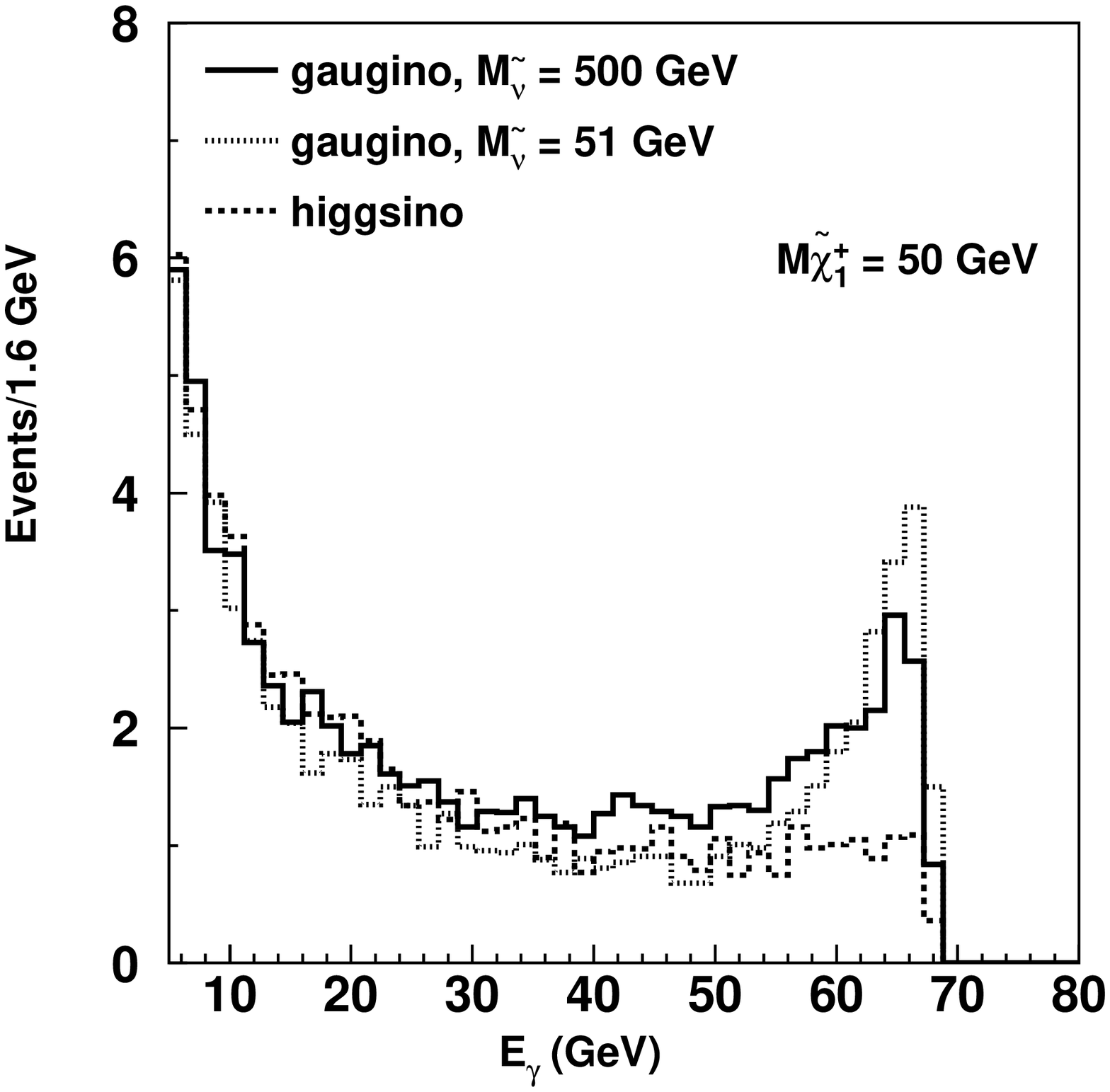}
\icaption{\label{ecuteff} ISR photon energy spectrum for 50 \GeV{} chargino 
pair production for a photon at least 10\dg{} away from the beam pipe. 
Distributions are shown
for gaugino-like chargino with $\m_{\snu} = 500 \GeV$ (solid line)
and $\m_{\snu} = \m_{\charg} + 1 \GeV$ (dotted line), and for 
higgsino-like charginos (dashed line) assuming an arbitrary 
common cross section
of about 6~pb.}
\end{center}
\end{figure}

The ISR energy spectrum in signal events, as shown in Figure
\ref{ecuteff}, depends on the relative contribution of the $s$-channel Z
exchange, which in turn depends on the ${\rm Z} \charg \charg$ coupling.
A softer photon spectrum
is expected in the case of higgsino-like charginos 
than in the case of gaugino-like charginos. In addition, when scalar neutrinos
are as light as charginos the $t$-channel diagram becomes important
and modifies the shape of the ISR energy spectrum 
as shown in Figure \ref{ecuteff}.

Chargino decay branching ratios in the low \dmns{} region change 
according to the opening of the various decay channels.
For $\dmns < 1.5 \GeV$, the hadronic decay channel in one or two pions 
is enhanced and the decay spectrum is more similar to 
$\tau^{\pm}$ decays than to W$^{\pm}$ decays.

The \charg{} decay length increases with decreasing \dmns.
For $\dmns \gtrsim 300 \MeV$, 
the \charg{} decays near the interaction vertex within 1 cm.
For $m_{\pi^{\pm}} < \dmns \lesssim 300 \MeV$, 
the \charg{} decays often inside the detector but a
few centimetres away from the interaction vertex.
For \dmns{} less than the $\pi^{\pm}$ mass, the \charg{} decays outside the detector
and appears as a heavy stable charged particle.

Chargino decays via virtual scalar lepton exchange occur
if scalar leptons are as light as charginos, but
only for gaugino-like charginos whose coupling 
to scalar leptons is strong enough.
In general, the chargino width is not affected by supersymmetric scalar
particles as long as they are more than 15 \GeV{} heavier 
than the \charg.
The same is true for chargino branching ratios if supersymmetric scalar
particles are at least 30 \GeV{} heavier than the \charg.

As an example, chargino decays are purely leptonic and the chargino decay length
is extremely short when the \snu{} is mass degenerate with
or lighter than the chargino. In the latter case, where
the two-body decay channel $\susy{\nu}_{\rm e} {\rm e}^\pm$ is open,
chargino decay lengths are smaller than 1 cm as long as \dmns{}
is more than the electron mass.

\section{Analysis}

\subsection{Preselection}

This preselection, as well as the following selections, are
tailored only on the expected signal distributions, because, as 
mentioned in Section \ref{simu}, the
simulation of Standard Model background processes is incomplete.

The aim of the preselection is to keep events with at least two charged 
particles not necessarily coming from the interaction vertex 
and an energetic photon.
The analysis relies mainly on the photon identification in the electromagnetic
calorimeter (BGO). In addition, we use the tracking chamber to detect
charged particles and to measure their momentum and eventually decay
length. Additional energy deposits 
due to the soft chargino decay products are allowed, but no strict
requirement is applied on them.
 
The BGO must contain an electromagnetic energy deposit more than 20\dg{} away 
from the beam pipe and with a transverse momentum compatible with equation 
(\ref{etcut}). There must be no track in a 1\dg{} sector in the $r - \phi$ plane
around the photon with a number of hits higher than 10 out of a
maximum of 62. This photon has also to be isolated in space:
neither a track with more than 10 hits nor
a significant energy deposit in the BGO should
be in a 15\dg{} cone around the  photon.
We also require that the event contains at least 2
tracks: one with at least 10 hits
and a distance to the interaction vertex in the $r - \phi$
plane (DCA) smaller than 1 cm; and a second track satisfying either
these criteria or having at least 20 hits.
For signal events satisfying these requirements, the trigger efficiency is 
nearly 100\%.

Since chargino decay products carry a small amount of energy,
the following selection criteria are applied:
no significant energy deposit in the low angle calorimeters 
covering $1.7^\circ \leq \theta \leq 9^\circ$;
less than 16 \GeV{} in the hadronic calorimeter;
less than 3 \GeV{} in the electromagnetic calorimeter 
between BGO barrel and endcaps; 
no muon track with a momentum greater than 10 \GeV.
Excluding the photon, the remaining energy in the BGO must be less than 
16 \GeV, and the total calorimetric energy 
must be less than 18 \GeV.
High multiplicity events are rejected by requiring 
less than 10 tracks and less than 15 BGO energy clusters.

\begin{figure}[p]
\begin{center}
\includegraphics[height=15cm]{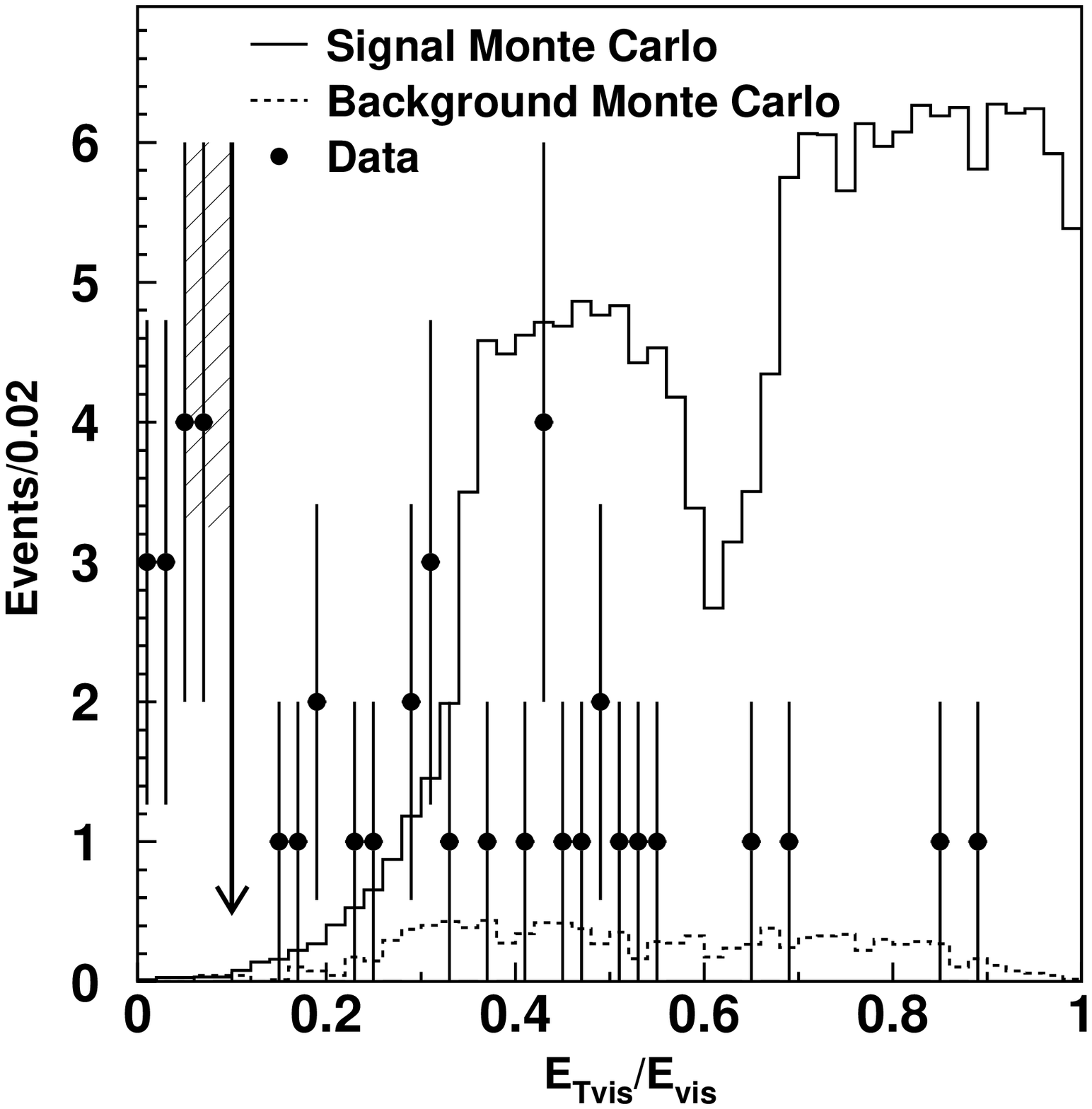}
\icaption{\label{etimb} Transverse energy imbalance for the data, the simulated
Standard Model background and the signal simulations  all masses and
\dmns{} folded in (arbitrary cross section of about 60 pb). The dip around 0.65
in the signal distribution is due to the gap at $\theta \sim 40\dg$
between BGO barrel and endcaps. The arrow shows the cut position.}
\end{center}
\end{figure}

After these cuts, 43 data events are selected for 10.8 expected from 
Standard Model processes which have been simulated.
Some of the remaining data events are not compatible with any
signals and they are probably due to photon-photon interactions.
This is illustrated in Figure \ref{etimb}, which shows the 
transverse energy imbalance ($\rm E_{Tvis}/E_{vis}$) 
distribution for the data, the simulated
Standard Model background and the signal (all masses 
and \dmns{} folded in the same distribution).  It is clear from this
distribution that data events with small transverse energy imbalance 
are not consistent with any
signal. Hence, we add the requirement that the transverse energy
imbalance must be greater than  0.1. 

The photon-photon interaction background should be
completely eliminated by the requirement of
equation (\ref{etcut}). In fact, this is true only for an absolutely
hermetic detector, such that deflected electrons are not missed.
To investigate this problem 
a sample of simulated $\ee \to \ee {\rm q \bar{q}}\gamma(\gamma)$
events, generated with {\tt PHOJET}\cite{PHOJET} for a mass of the 
$\gamma\gamma$ system greater than 3 \GeV, is used.
The expected background from this source, concentrated at low values
of transverse energy imbalance, is 0.8 events.
This result is not conclusive because masses of the 
$\gamma\gamma$ system below 3 \GeV{} can not be simulated due to
the aforementioned uncertainties on the photon-photon interaction 
cross section. Nevertheless, it shows that a non-zero background from
photon-photon interaction is expected, due to 
small residual inefficiencies to tag electrons at 
low angles ($\theta_d < 10\dg$), which might explain
the excess of data.

Finally, 29 data events are selected and 10.7 are expected from the simulated Standard
Model background (4.2 from $\mu^+\mu^-\gam(\gam)$, 5.5 from
$\tau^+\tau^-\gam(\gam)$, $<0.06$ from $\rm e^+e^-\gam(\gam)$, 
0.8 from \nngg, 0.2 from W$^+$W$^-$$\gamma(\gam)$).

\subsection{Selections}

Three different selections are devised according to the \dm range explored:
low \dmns, very low \dmns{} and ultra low \dmns{} selections.

The low \dmns{} selection is optimized for \dmns{} around 3 \GeV. At such \dmns,
charginos decay promptly and their decay products carry enough energy to reach
the BGO. This selection requires that the event contains at least
2 tracks with a DCA smaller than 1~cm and a second energy deposit in the BGO.

The very low \dmns{} selection is optimized for \dmns{} around 1 \GeV{} 
whereas the ultra low \dmns{} selection is optimized 
for \dmns{} around 300 \MeV{} and less. 
At such small \dmns, no muons are able to reach the muon chamber. Therefore,
events must not contain muon tracks. 
The other requirements on the calorimetric energy are modified
according to the smaller energy deposited by the chargino decay products.
No requirements on the track momentum are applied in the 
ultra low \dmns{} selection to take into account the possible high 
momentum tracks produced by long lived charginos. Table \ref{cuts}
lists all selection cuts for all \dm regions.
\begin{table}[h]
\begin{center} 
\begin{tabular}{|cc|c|c|c|}
\cline{3-5}
\multicolumn{2}{c}{} & \multicolumn{3}{|c|}{Selection} \\
\hline
Cut && low \dmns & very low \dmns &  ultra low \dmns  \\
\hline 
hadronic calorimeter energy &$<$& 12 \GeV & 10 \GeV & 10 \GeV \\
\hline
${\rm E}_{\rm BGO} - {\rm E}_{\gamma}$ &$<$ & 10 \GeV & 6 \GeV & 1 \GeV \\
\hline
remaining calorimetric energy &$<$& 12 \GeV & 8 \GeV & 6 \GeV \\
\hline
muon momentum &$<$& 8 \GeV & No muon & No muon \\
\hline
P$_{\rm t}$ track &$<$& 10 \GeV & 4 \GeV & none \\
\hline
transverse energy imbalance &$>$& 0.1 & 0.2 & 0.3 \\
\hline
longitudinal energy imbalance &$<$& 0.85 & none & none \\
\hline
number of tracks &$<$& 10 & 7 & 5 \\
\hline 
number of BGO energy clusters &$<$& 15 & 10 & 6 \\
\hline
isolation angle of the photon &$<$& 160\dg{} & none & none \\ 
\hline
\end{tabular} 
\icaption{\label{cuts}Requirements for all selections.}\\
\end{center}
\end{table}

The number of selected data events and expected backgrounds are shown in Table 
\ref{nev}. This table displays also the number of events selected by 
the three selections combined.
\begin{table}[h]
\begin{center} 
\begin{tabular}{|c|c|c|c|c|}
\cline{2-5}
\multicolumn{1}{c}{} & \multicolumn{4}{|c|}{Selection} \\
\cline{2-5}
\multicolumn{1}{c|}{} & low \dmns & very low \dmns & ultra low \dmns & Combined \\
\hline
$\mu^+\mu^-\gam(\gam)$ & 0.41 & 0.08 & 0.63 & 0.94 \\
\hline
$\tau^+\tau^-\gam(\gam)$ & 1.34 & 0.39 & 0.08 & 1.44 \\
\hline
\nngg & 0.54 & 0.57 & 0.20 & 0.64 \\
\hline
W$^+$W$^-$$\gamma(\gam)$ & 0.03 & 0.01 & 0. & 0.03 \\
\hline
\hline
Total & 2.32 & 1.05 & 0.91 & 3.05\\
\hline
Data & 6 & 1 & 1 & 8 \\
\hline
\end{tabular} \\
\icaption{\label{nev} Number of selected data events and expected  
from simulated Standard Model backgrounds. The last column displays the same
numbers for the three selections combined.}
\end{center}
\end{table} 
Finally, for a given \charg{} mass, events are considered as candidates
only if the effective centre-of-mass energy is high enough, i.e.
the photon recoil mass is greater than twice the \charg{} mass.

\section{Results}

\begin{figure}[p]
\begin{center}
\includegraphics[height=15cm]{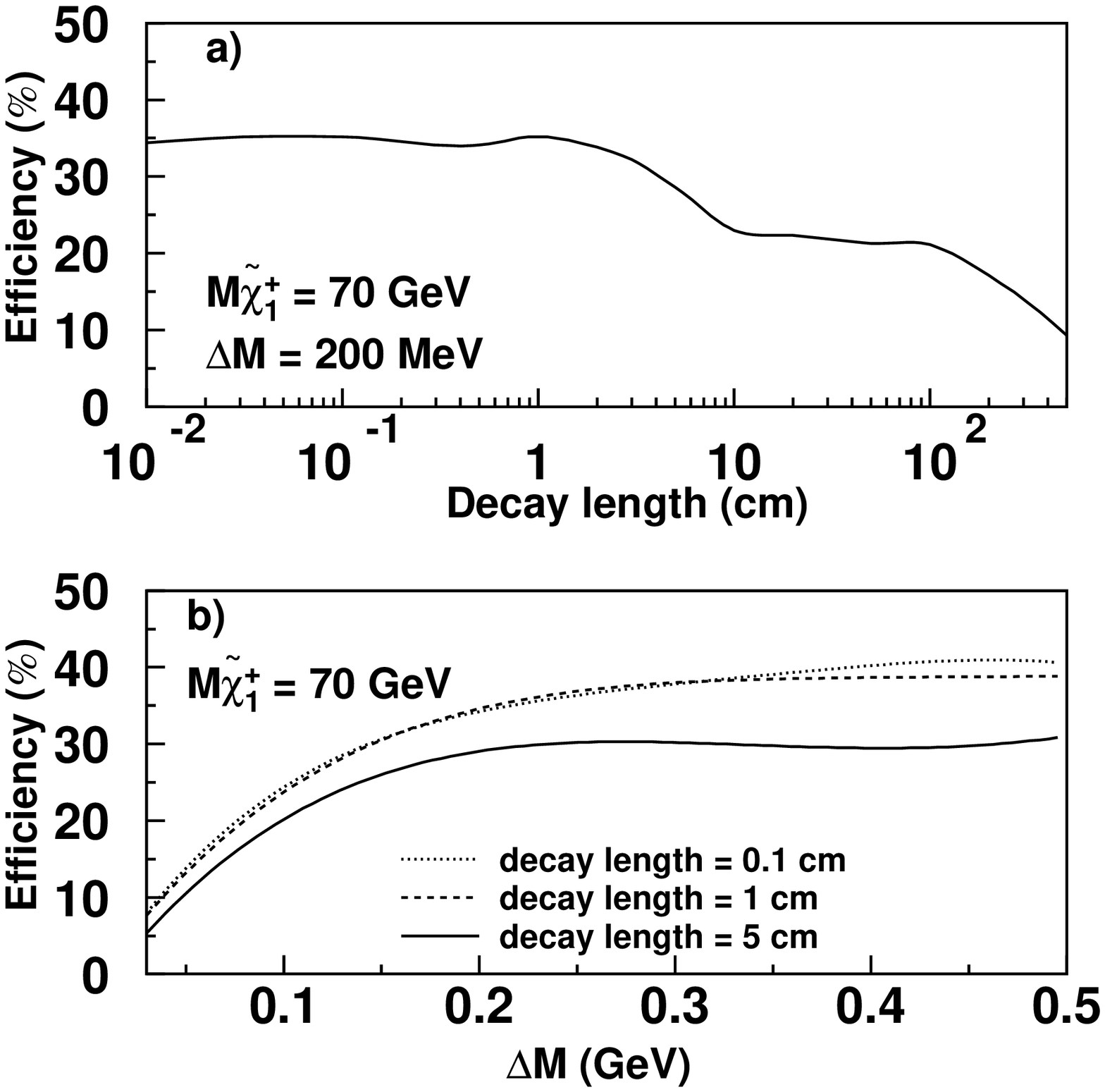}
\icaption{\label{brdm} a) Selection efficiency as a function of the chargino decay
  length for $\dmns = 200 \MeV$. b) Selection efficiency as a
  function of \dmns{} for several values of the decay length.
  Both plots are for a 70 \GeV{} chargino
  and for an ISR photon within fiducial cuts.}
\end{center}
\end{figure}

The selection efficiency for short-lived charginos with a photon within 
fiducial cuts is about 35\% for $\dmns \geq 200 \MeV$.
In Figure \ref{brdm}a the selection efficiency is shown as a function
of the chargino decay length. For decay lengths of tens of centimetres
the efficiency decreases to approximately 20\%, because in that range
highly ionising chargino tracks reach the tracking chamber and the
BGO calorimeter. The efficiency drops for decay lengths of
a few metres.  This drop is due to long-lived
charginos which produce high momentum tracks in the muon chambers and
events without missing energy. This kind of signal, which suffers from the 
$\mu^+ \mu^- \gamma(\gamma)$ background, is taken into account by the
search for stable heavy charged particles \cite{heavystab}.

In Figure \ref{brdm}b is shown the selection efficiency as a
function of \dmns{} for several values of the chargino decay length.
The drop for $\dmns < 100 \MeV$ is due to the magnetic field which
bends low momentum tracks, such that they do not reach the tracking
chamber.
This is a major experimental limitation in the search of short-lived
charginos for $\dmns < 50 \MeV$. In particular, 
for $\susy{\nu}_{\rm e}$ or $\susy{e}$ mass degenerate with charginos 
the decay length is extremely short 
as long as \dmns{} is above the electron mass.

\begin{figure}[p]
\begin{center}
\includegraphics[height=15cm]{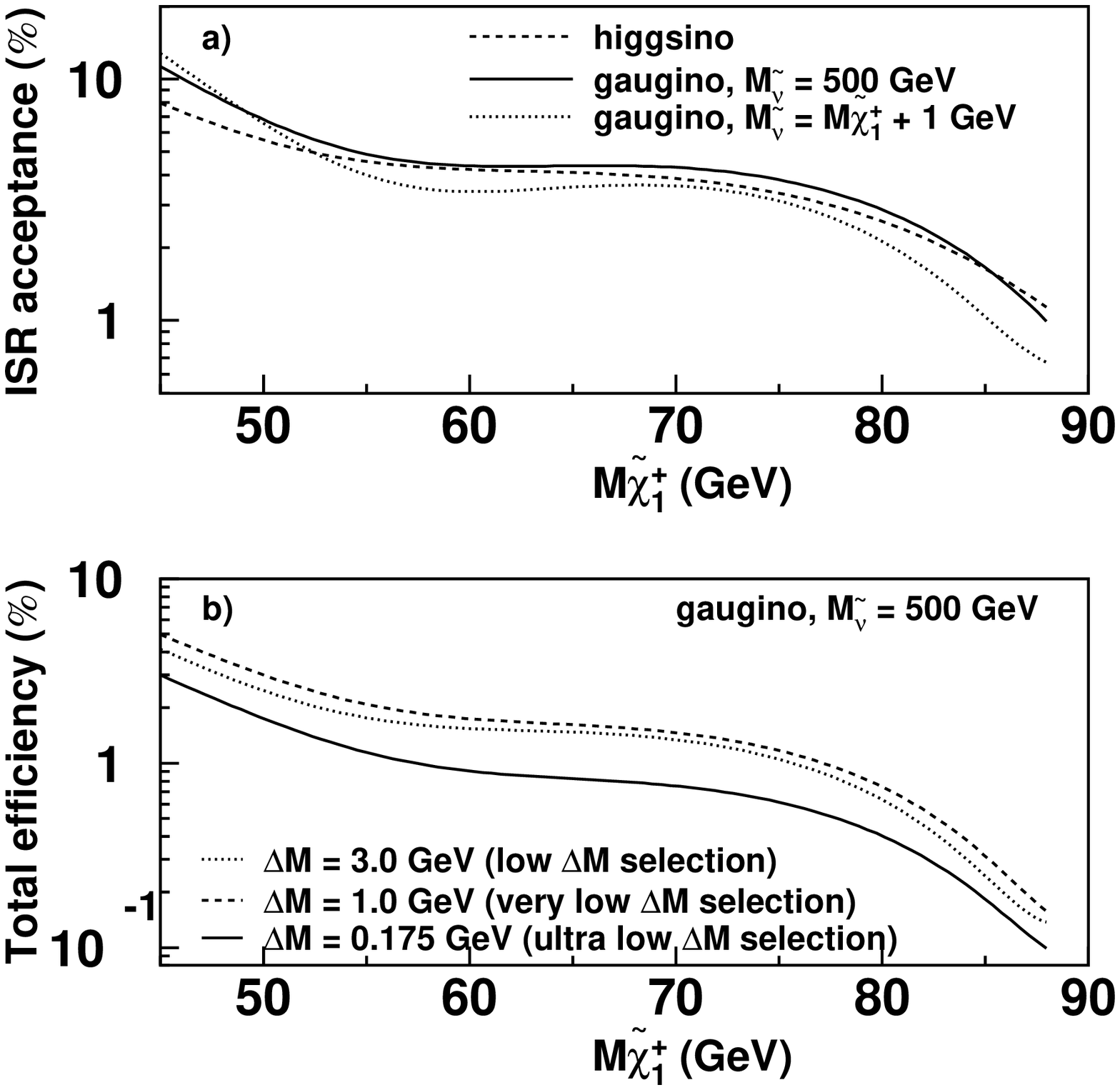}
\icaption{\label{effic} a) Acceptance for events with 
an ISR photon within fiducial cuts
as a function of $\m_{\charg}$ for several chargino mixtures and \snu{} masses.
b) Total selection efficiency as a function of $\m_{\charg}$ 
for several \dmns{} values and in the case of a gaugino-like \charg
and a heavy \snu.}
\end{center}
\end{figure}

In Figure \ref{effic}a is shown the acceptance for events with 
an ISR photon within
fiducial cuts as a function of the chargino mass.
This acceptance also depends on the chargino mixture and on
the \snu{} mass. The total efficiency, product of the efficiency in Figure
\ref{brdm} and the acceptance in Figure \ref{effic}a, is derived as shown 
in Figure \ref{effic}b
for a gaugino-like \charg.
In the same way, efficiencies 
are also estimated for a higgsino-like \charg, and for a gaugino-like \charg{}
in the light \snu{} case. 
The evolution of the efficiency with $\m_{\charg}$ is essentially 
governed by the ISR spectrum.

A Standard Model process with a similar signature to the one
we search in this analysis is \nngg. 
The analysis of \nngg{} events has already been
performed on 189 \GeV{} data \cite{papnng}, resulting in a good agreement
between observation and Standard Model expectation.
Systematic uncertainties affecting the photon identification, which
are smaller than 1\%, are relevant for the systematics on
the efficiency shown in Figure \ref{effic}b.
The loose requirements on the soft charged tracks 
induce negligible systematics.
Larger  uncertainties are due to
Monte Carlo statistical errors. They range between 6\% and 7\%
according to the chargino mixture and decay length.
These systematics are neglected hereafter since the derived
limits are already conservative due to the omission
of the photon-photon interaction processes, resulting in an underestimation 
of the background.

\begin{figure}[p]
\begin{center}
\includegraphics[height=15cm]{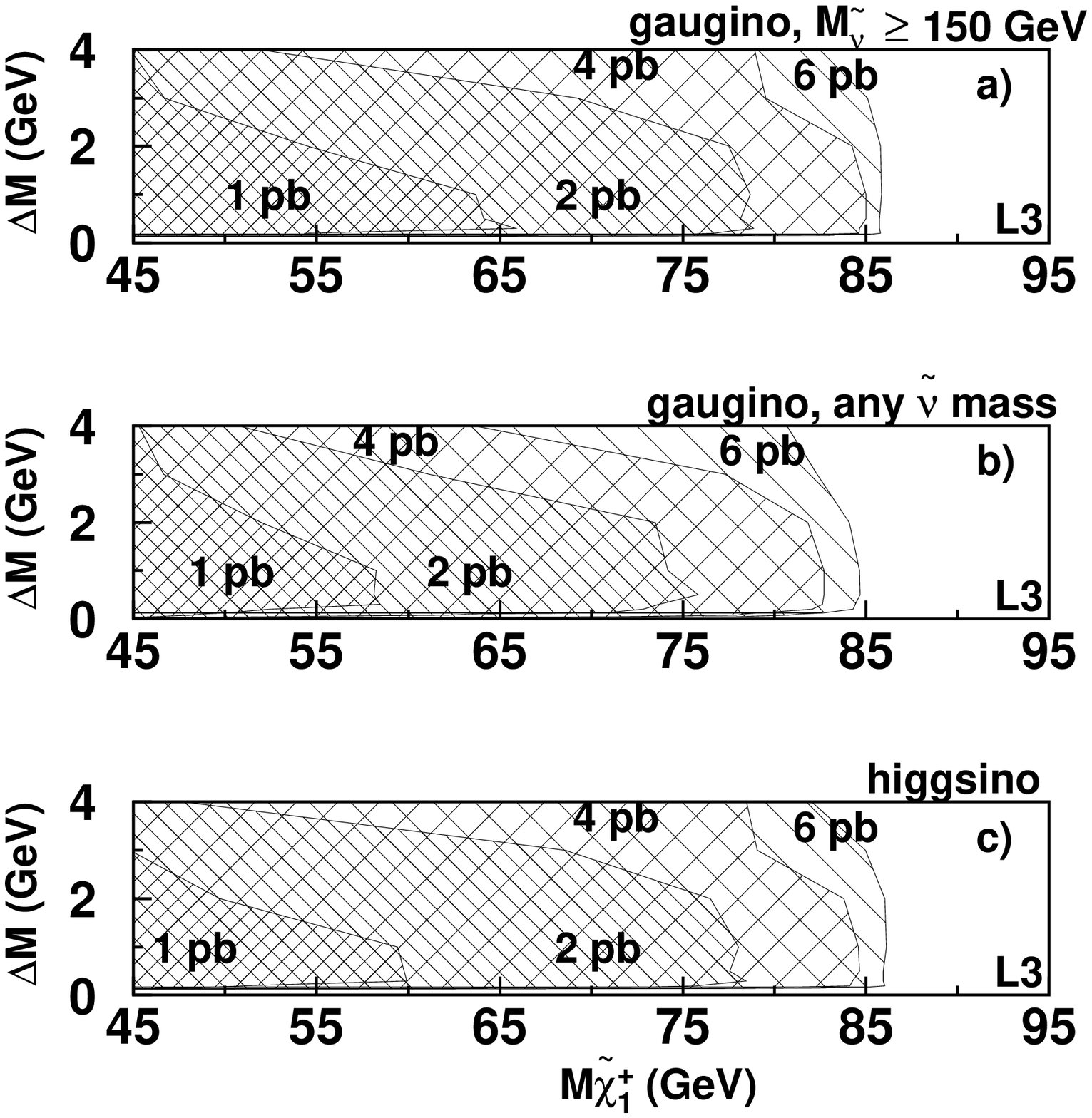}
\icaption{\label{siglim} 95\% C.L. cross section upper limits 
 as a function of $\m_{\charg}$ and \dmns{} for gaugino-like
 charginos with a) $ \m_{\snu} \gtrsim 150 \GeV$ and b) any \snu{} mass, 
and c) for higgsino-like charginos. For these upper limits, for any \dm the 
smallest decay length compatible with the MSSM is used.}
\end{center}
\end{figure}

Overall we select 8 data events for 3.1 expected from the simulated
Standard Model processes. We use
the signal efficiency, the background prediction and the number of selected events, 
 to derive a  95\% confidence level upper limit on the chargino pair-production
cross section as a function of \dmns{} and $\m_{\charg}$ as shown in 
Figure \ref{siglim}. Those limits are obtained by combining the three selections,
according to the method in Ref. \cite{stat} modified
to include the background subtraction.

\section{Interpretations in the MSSM}

In the MSSM, chargino and neutralino masses depend on 4 parameters: 
$M_1$ the U(1) gaugino mass, $M_2$ the SU(2) gaugino mass, $\mu$ the Higgs
mixing parameter and $\tanb$ the ratio of the two Higgs vacuum 
expectation values. 

Generally, equal gaugino masses at the GUT scale are assumed, such
that $M_1$ and $M_2$ are not independent parameters.
In this case, low \dmns{} values are possible only if $| \mu | \muchless M_2$ and for
$M_2$ values larger than a few \TeV.
However, models with gaugino mass non-universality can be considered \cite{snow}
and as mentioned in Ref. \cite{gsearch} low \dmns{} regions become 
more natural in these models \cite{gstring}.

Once the unification relation is relaxed, three main regions can lead
to low \dmns: $| \mu | \gg M_2$; in this region the \charg{} mass is 
almost equal to $M_2$, \charg{} and \neut{} are both gaugino-like
 and \dmns{} can be small if $M_1 \gtrsim M_2$.
$| \mu | \muchless M_2$; in this region \neut{} and \charg{} are 
both higgsino-like and nearly mass degenerate, independently
of the values of $M_1$ and $M_2$. Their masses are almost equal to
$| \mu |$.
$M_1 \gtrsim 4 M_2$; in this region of the parameter space
\neut{} and \charg{} mass degeneracy can be obtained for pure or 
mixed \charg{} states.

\begin{figure}[p]
\begin{center}
\includegraphics[height=15cm]{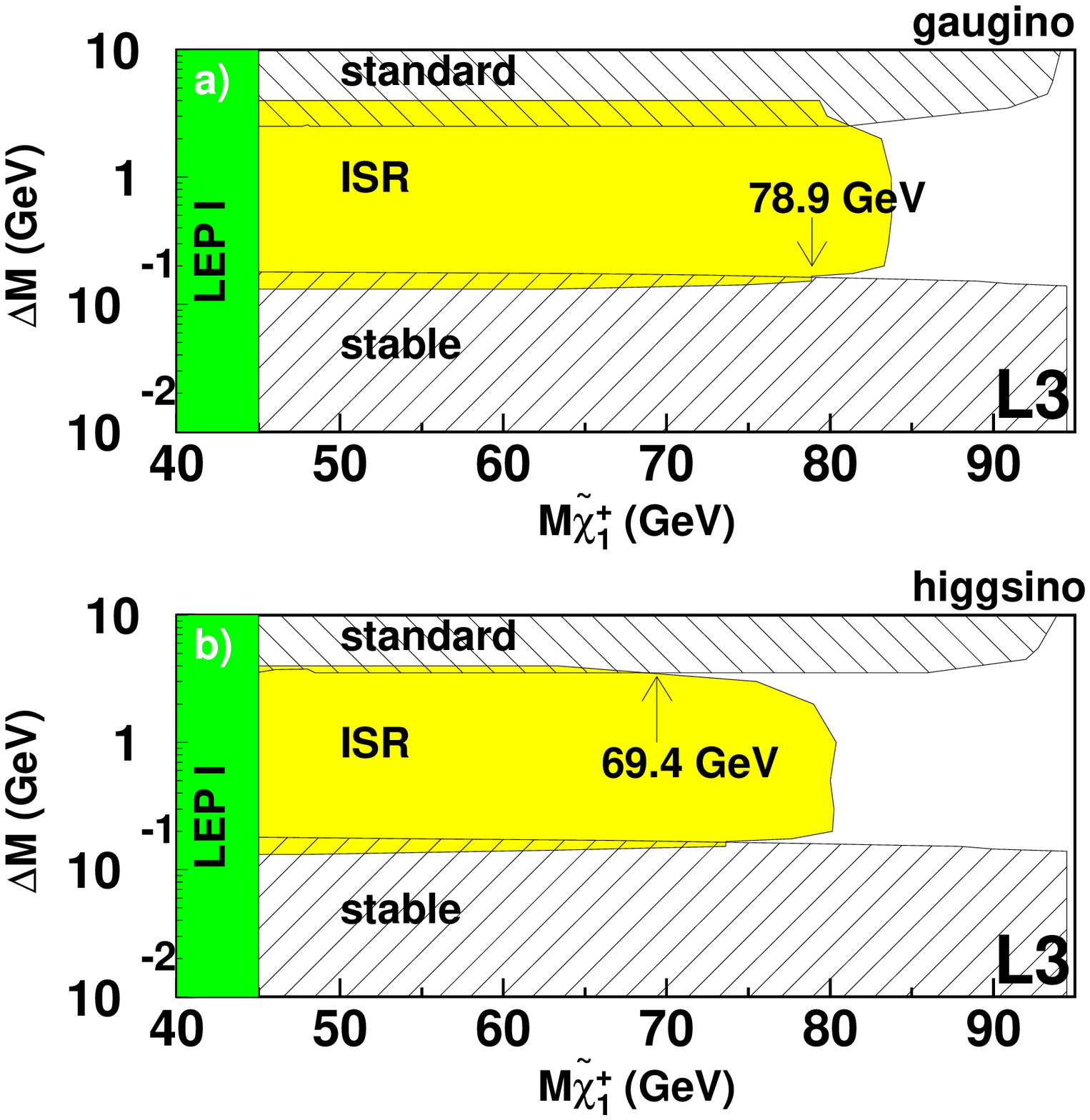}
\caption{\label{exclmix} 95\% C.L. chargino mass limits as a function of \dmns{}
a) for a gaugino-like \charg{} with a heavy \snu{} and b) for higgsino-like
 \charg{}.
On each plot the top hatched area corresponds to the standard search\protect\cite{stsearch}, 
the bottom hatched area corresponds to the search for heavy stable charged 
particles\protect\cite{heavystab} and the grey shaded area to this search.
The exclusion from LEP1 results on the Z width \protect\cite{caso} is
also shown.}
\end{center}
\end{figure}

\begin{figure}[p]
\begin{center}
\includegraphics[height=15cm]{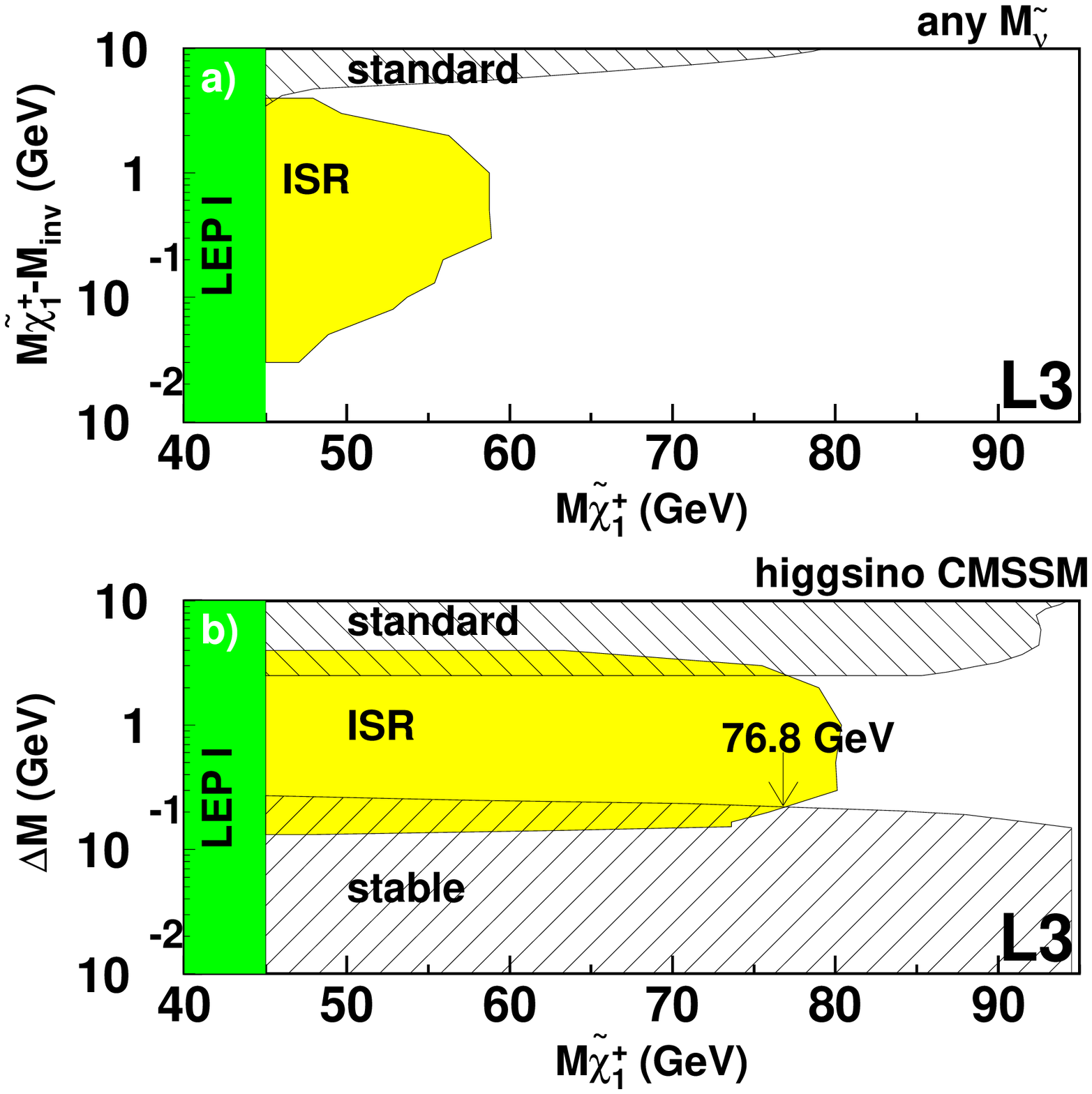}
\caption{\label{exclmix2} 95\% C.L. chargino mass limits 
a) as a function of $\m_{\charg}-\m_{inv}$
for a gaugino-like \charg{} and for any \snu{} mass and b) as a function of \dm
for a higgsino-like \charg{} in the Constrained MSSM (CMSSM).
On each plot the top hatched area corresponds to the standard search\protect\cite{stsearch}, 
the bottom hatched area corresponds to the search for heavy stable charged 
particles\protect\cite{heavystab} and the grey shaded area to this
search. The exclusion from LEP1 results on the Z width \protect\cite{caso} is
also shown.}
\end{center}
\end{figure}

We derive chargino mass limits as a function of \dmns{} for all these
scenarios. The limits are shown in Figures \ref{exclmix} and \ref{exclmix2}. 
For \dmns{} between 300 \MeV{} and 1 \GeV, 
gaugino-like \charg{} are excluded up to 
83.5 \GeV{} if the \snu{} is heavy (Figure \ref{exclmix}a) 
and up to 58.8 \GeV{} for any \snu{} mass (Figure \ref{exclmix2}a), 
where cross sections  smaller by an order of magnitude are predicted.
In Figure \ref{exclmix2}a the limit is shown as a function of 
$\m_{\charg} - \m_{\rm inv}$, where the invisible particle can
be either the lightest neutralino or the scalar neutrino.
For the same \dmns{} range, higgsino-like \charg{} are excluded up to 80.0
\GeV{} (Figure \ref{exclmix}b).

The results of this search are combined with the results of the
standard \charg{} search\cite{stsearch}
and of the stable heavy charged particles search\cite{heavystab} as shown 
in Figures \ref{exclmix} and  \ref{exclmix2}.
The search with an ISR photon fills the gap between the two previous analyses 
and allows to derive direct search \charg{} mass limits independent of
\dmns.
 
The intersection between the ISR search and the stable heavy charged
particles search occurs for decay lengths of order 10 cm. However,
the relation between the chargino decay length and the \dmns{} is
dependent on the supersymmetric parameters. 
Due to the mild dependence of the efficiency on \dmns{}, a scan on 
the MSSM parameters is done to check the size of the overlap between 
these two searches. The relation between the decay length and \dmns{} used to
derive the limits shown in Figures \ref{exclmix}a, \ref{exclmix}b
and \ref{exclmix2}a, is always the most conservative independently
of the choice for the chargino mixture used for the cross section
calculation. On the contrary, in Figure \ref{exclmix2}b, where the
Constrained MSSM \cite{MSSM_GUT} is used, chargino and neutralino
mixtures are uniquely defined and used both for the decay length and
cross section calculations.

Assuming that scalar particles are sufficiently heavy to be able to neglect their 
contributions in chargino production and decay, the following chargino mass limits
are derived at 95\% C.L.:
\begin{center} 
\begin{tabular}{cc}
$\m_{\charg} > 78.9 \GeV{}$ & gaugino-like charginos \\
$\m_{\charg} > 69.4 \GeV{}$ & higgsino-like charginos.
\end{tabular}
\end{center}
As mentioned in Section \ref{pheno}, the chargino decay length also
depends on the masses of the scalar particles. For light scalar quarks,
hadronic decays are enhanced and a sharp change from short to
long-lived charginos is expected for \dmns{} around the pion mass.
In this case, the chargino mass limit is 0.5 \GeV{} lower 
for gaugino-like \charg, while it is unchanged for higgsino-like \charg.
In the case of light scalar taus or $\susy{\nu}_{\tau}$, the chargino 
mass limit is unchanged for gaugino-like \charg{} and lowered by 0.8 \GeV{} for
higgsino-like \charg.
For light scalar muons or $\susy{\nu}_{\mu}$ 
the chargino mass limit is lowered by 1.3 \GeV{} for gaugino-like \charg, 
while it is unchanged for higgsino-like \charg.

No direct search mass limits are derived for light
$\susy{\nu}_{\rm e}$ or $\susy{e}$ masses, 
because there is still an uncovered \dmns{} region between
the stable heavy charged particles search and the search with an ISR
photon. This gap
is due to the shorter \charg{} decay length when these scalar particles 
are mass degenerate with or lighter than the chargino.
In such a case, charginos are stable only for \dmns{} below the electron
mass.

Figure \ref{exclmix2}b shows the exclusion in the
Constrained MSSM \cite{MSSM_GUT} in the  higgsino region (very large $M_2$ values).
The exclusion from the standard search is obtained by combining 
the contribution of the $\ee \to \susy{\chi}_2^0 \neut$ process\cite{stsearch}
 with the chargino pair production:
\begin{center} 
\begin{tabular}{cc}
$\m_{\charg} > 76.8 \GeV{}$ & Constrained MSSM.
\end{tabular}
\end{center}
In the Constrained MSSM, by using the scalar lepton search \cite{slepton}, 
a limit on the \charg{} mass of 67.7 \GeV{} is also derived in the 
light \snu{} case, assuming no mixing in the scalar tau sector.
This limit of 67.7 \GeV{} is an absolute lower limit for the
chargino in the Constrained MSSM parameter space, since
the very large $M_2$ domain is now excluded by this search.


\section*{Acknowledgments}

We  express our gratitude to the CERN accelerator divisions for the
excellent performance of the LEP machine. We also acknowledge
and appreciate the effort of the engineers, technicians and support staff 
who have participated in the construction and maintenance of this experiment.

%
%
\newpage
\section*{Author List}
\typeout{   }     
\typeout{Using author list for paper 200 -?}
\typeout{$Modified: Mon Jan 10 09:41:54 2000 by clare $}
\typeout{!!!!  This should only be used with document option a4p!!!!}
\typeout{   }
%
%
%
%
%
%

\newcount\tutecount  \tutecount=0
\def\tutenum#1{\global\advance\tutecount by 1 \xdef#1{\the\tutecount}}
\def\tute#1{$^{#1}$}
\tutenum\aachen            
\tutenum\nikhef            
\tutenum\mich              
\tutenum\lapp              
\tutenum\basel             
\tutenum\lsu               
\tutenum\beijing           
\tutenum\berlin            
\tutenum\bologna           
\tutenum\tata              
\tutenum\ne                
\tutenum\bucharest         
\tutenum\budapest          
\tutenum\mit               
\tutenum\debrecen          
\tutenum\florence          
\tutenum\cern              
\tutenum\wl                
\tutenum\geneva            
\tutenum\hefei             
\tutenum\seft              
\tutenum\lausanne          
\tutenum\lecce             
\tutenum\lyon              
\tutenum\madrid            
\tutenum\milan             
\tutenum\moscow            
\tutenum\naples            
\tutenum\cyprus            
\tutenum\nymegen           
\tutenum\caltech           
\tutenum\perugia           
\tutenum\cmu               
\tutenum\prince            
\tutenum\rome              
\tutenum\peters            
\tutenum\potenza           
\tutenum\salerno           
\tutenum\ucsd              
\tutenum\santiago          
\tutenum\sofia             
\tutenum\korea             
\tutenum\alabama           
\tutenum\utrecht           
\tutenum\purdue            
\tutenum\psinst            
\tutenum\zeuthen           
\tutenum\eth               
\tutenum\hamburg           
\tutenum\taiwan            
\tutenum\tsinghua          
{
\parskip=0pt
\noindent
{\bf The L3 Collaboration:}
\ifx\selectfont\undefined
 \baselineskip=10.8pt
 \baselineskip\baselinestretch\baselineskip
 \normalbaselineskip\baselineskip
 \ixpt
\else
 \fontsize{9}{10.8pt}\selectfont
\fi
\medskip
\tolerance=10000
\hbadness=5000
\raggedright
\hsize=162truemm\hoffset=0mm
\def\r{\rlap,}
\noindent

M.Acciarri\r\tute\milan\
P.Achard\r\tute\geneva\ 
O.Adriani\r\tute{\florence}\ 
M.Aguilar-Benitez\r\tute\madrid\ 
J.Alcaraz\r\tute\madrid\ 
G.Alemanni\r\tute\lausanne\
J.Allaby\r\tute\cern\
A.Aloisio\r\tute\naples\ 
M.G.Alviggi\r\tute\naples\
G.Ambrosi\r\tute\geneva\
% ...
\newpage
\begin{list}{A}{\itemsep=0pt plus 0pt minus 0pt\parsep=0pt plus 0pt minus 0pt
                \topsep=0pt plus 0pt minus 0pt}
\item[\aachen]
 I. Physikalisches Institut, RWTH, D-52056 Aachen, FRG$^{\S}$\\
 III. Physikalisches Institut, RWTH, D-52056 Aachen, FRG$^{\S}$
\item[\nikhef] National Institute for High Energy Physics, NIKHEF, 
     and University of Amsterdam, NL-1009 DB Amsterdam, The Netherlands
\item[\mich] University of Michigan, Ann Arbor, MI 48109, USA
\item[\lapp] Laboratoire d'Annecy-le-Vieux de Physique des Particules, 
     LAPP,IN2P3-CNRS, BP 110, F-74941 Annecy-le-Vieux CEDEX, France
\item[\basel] Institute of Physics, University of Basel, CH-4056 Basel,
     Switzerland
\item[\lsu] Louisiana State University, Baton Rouge, LA 70803, USA
\item[\beijing] Institute of High Energy Physics, IHEP, 
  100039 Beijing, China$^{\triangle}$ 
\item[\berlin] Humboldt University, D-10099 Berlin, FRG$^{\S}$
\item[\bologna] University of Bologna and INFN-Sezione di Bologna, 
     I-40126 Bologna, Italy
\item[\tata] Tata Institute of Fundamental Research, Bombay 400 005, India
\item[\ne] Northeastern University, Boston, MA 02115, USA
\item[\bucharest] Institute of Atomic Physics and University of Bucharest,
     R-76900 Bucharest, Romania
\item[\budapest] Central Research Institute for Physics of the 
     Hungarian Academy of Sciences, H-1525 Budapest 114, Hungary$^{\ddag}$
\item[\mit] Massachusetts Institute of Technology, Cambridge, MA 02139, USA
\item[\debrecen] KLTE-ATOMKI, H-4010 Debrecen, Hungary$^\P$
\item[\florence] INFN Sezione di Firenze and University of Florence, 
     I-50125 Florence, Italy
\item[\cern] European Laboratory for Particle Physics, CERN, 
     CH-1211 Geneva 23, Switzerland
\item[\wl] World Laboratory, FBLJA  Project, CH-1211 Geneva 23, Switzerland
\item[\geneva] University of Geneva, CH-1211 Geneva 4, Switzerland
\item[\hefei] Chinese University of Science and Technology, USTC,
      Hefei, Anhui 230 029, China$^{\triangle}$
\item[\seft] SEFT, Research Institute for High Energy Physics, P.O. Box 9,
      SF-00014 Helsinki, Finland
\item[\lausanne] University of Lausanne, CH-1015 Lausanne, Switzerland
\item[\lecce] INFN-Sezione di Lecce and Universit\'a Degli Studi di Lecce,
     I-73100 Lecce, Italy
\item[\lyon] Institut de Physique Nucl\'eaire de Lyon, 
     IN2P3-CNRS,Universit\'e Claude Bernard, 
     F-69622 Villeurbanne, France
\item[\madrid] Centro de Investigaciones Energ{\'e}ticas, 
     Medioambientales y Tecnolog{\'\i}cas, CIEMAT, E-28040 Madrid,
     Spain${\flat}$ 
\item[\milan] INFN-Sezione di Milano, I-20133 Milan, Italy
\item[\moscow] Institute of Theoretical and Experimental Physics, ITEP, 
     Moscow, Russia
\item[\naples] INFN-Sezione di Napoli and University of Naples, 
     I-80125 Naples, Italy
\item[\cyprus] Department of Natural Sciences, University of Cyprus,
     Nicosia, Cyprus
\item[\nymegen] University of Nijmegen and NIKHEF, 
     NL-6525 ED Nijmegen, The Netherlands
\item[\caltech] California Institute of Technology, Pasadena, CA 91125, USA
\item[\perugia] INFN-Sezione di Perugia and Universit\'a Degli 
     Studi di Perugia, I-06100 Perugia, Italy   
\item[\cmu] Carnegie Mellon University, Pittsburgh, PA 15213, USA
\item[\prince] Princeton University, Princeton, NJ 08544, USA
\item[\rome] INFN-Sezione di Roma and University of Rome, ``La Sapienza",
     I-00185 Rome, Italy
\item[\peters] Nuclear Physics Institute, St. Petersburg, Russia
\item[\potenza] INFN-Sezione di Napoli and University of Potenza, 
     I-85100 Potenza, Italy
\item[\salerno] University and INFN, Salerno, I-84100 Salerno, Italy
\item[\ucsd] University of California, San Diego, CA 92093, USA
\item[\santiago] Dept. de Fisica de Particulas Elementales, Univ. de Santiago,
     E-15706 Santiago de Compostela, Spain
\item[\sofia] Bulgarian Academy of Sciences, Central Lab.~of 
     Mechatronics and Instrumentation, BU-1113 Sofia, Bulgaria
\item[\korea] Center for High Energy Physics, Adv.~Inst.~of Sciences
     and Technology, 305-701 Taejon,~Republic~of~{Korea}
\item[\alabama] University of Alabama, Tuscaloosa, AL 35486, USA
\item[\utrecht] Utrecht University and NIKHEF, NL-3584 CB Utrecht, 
     The Netherlands
\item[\purdue] Purdue University, West Lafayette, IN 47907, USA
\item[\psinst] Paul Scherrer Institut, PSI, CH-5232 Villigen, Switzerland
\item[\zeuthen] DESY, D-15738 Zeuthen, 
     FRG
\item[\eth] Eidgen\"ossische Technische Hochschule, ETH Z\"urich,
     CH-8093 Z\"urich, Switzerland
\item[\hamburg] University of Hamburg, D-22761 Hamburg, FRG
\item[\taiwan] National Central University, Chung-Li, Taiwan, China
\item[\tsinghua] Department of Physics, National Tsing Hua University,
      Taiwan, China
\item[\S]  Supported by the German Bundesministerium 
        f\"ur Bildung, Wissenschaft, Forschung und Technologie
\item[\ddag] Supported by the Hungarian OTKA fund under contract
numbers T019181, F023259 and T024011.
\item[\P] Also supported by the Hungarian OTKA fund under contract
  numbers T22238 and T026178.
\item[$\flat$] Supported also by the Comisi\'on Interministerial de Ciencia y 
        Tecnolog{\'\i}a.
\item[$\sharp$] Also supported by CONICET and Universidad Nacional de La Plata,
        CC 67, 1900 La Plata, Argentina.
\item[$\diamondsuit$] Also supported by Panjab University, Chandigarh-160014, 
        India.
\item[$\triangle$] Supported by the National Natural Science
  Foundation of China.
\item[\dag] Deceased.
\end{list}
}
\vfill






\newpage
%
\bibliography{paper}

\begin{thebibliography}{10}

\bibitem{mssm}
H.P. Nilles, Phys. Rep. {\bf 110} (1984) 1;\\ H.E. Haber and G.L. Kane, Phys.
  Rep. {\bf 117} (1985) 75.

\bibitem{stsearch}
L3 Collaboration, M. Acciarri \etal, CERN-EP/99-127, accepted by Phys. Lett. B.

\bibitem{mark2}
K. Riles \etal, Phys. Rev. {\bf D 42} (1990) 1.

\bibitem{gsearch}
C.H. Chen, M. Drees and J.F. Gunion,
\newblock  Phys. Rev. Lett. {\bf 76}  (1996) 2002,
\newblock  Erratum Phys. Rev. Lett. {\bf 82} (1999) 3192.

\bibitem{DELPHI183}
DELPHI Collaboration, P. Abreu et al.,
\newblock  Eur. Phys. J. {\bf C 11}  (1999) 1--17.

\bibitem{L3-DETECTOR}
L3 Collaboration, B. Adeva \etal, Nucl. Instr. and Meth. {\bf A 289} (1990) 35;
  \\ M. Chemarin \etal, Nucl. Instr. and Meth. {\bf A 349} (1994) 345; \\ M.
  Acciarri \etal, Nucl. Instr. and Meth. {\bf A 351} (1994) 300; \\ G. Basti
  \etal, Nucl. Instr. and Meth. {\bf A 374} (1996) 293; \\ I.C. Brock \etal,
  Nucl. Instr. and Meth. {\bf A 381} (1996) 236; \\ A. Adam \etal, Nucl. Instr.
  and Meth. {\bf A 383} (1996) 342.

\bibitem{KORALZ}
The {\tt KORALZ} version 4.02 is used.\\ S. Jadach, B.F.L. Ward and Z. W\c{a}s,
  Comp. Phys. Comm. {\bf 79} (1994) 503.

\bibitem{KORALW}
{\tt KORALW} version 1.33 is used.\\ M.~Skrzypek \etal, \CPC {\bf 94} (1996)
  216;\\ M.~Skrzypek \etal, \PL {\bf B 372} (1996) 289.

\bibitem{TEEGG}
The {\tt TEEGG} version 7.1 is used.\\ D. Karlen, Nucl. Phys. {\bf B 289}
  (1987) 23.

\bibitem{susygen2.2}
{\tt SUSYGEN} version 2.2 is used.\\ S. Katsanevas and P. Morawitz, \CPC {\bf
  112} (1998) 227.

\bibitem{geant}
The L3 detector simulation is based on GEANT Version 3.15.\\ See R. Brun et
  al., ``GEANT 3'', CERN DD/EE/84-1 (Revised), September 1987.

\bibitem{gheisha}
The {\tt GHEISHA} program,
\newblock  H. Fesefeldt, RWTH Aachen Report PITHA 85/02 (1985).

\bibitem{PHOJET}
{\tt PHOJET} version 1.10 is used. \\ R.~Engel, Z. Phys. C 66 (1995) 203; \\
  R.~Engel and J.~Ranft, Phys. Rev. D 54 (1996) 4244.

\bibitem{heavystab}
L3 Collaboration, M. Acciarri \etal, Phys. Lett. {\bf B 462} (1999) 354.

\bibitem{papnng}
L3 Collaboration, M. Acciarri \etal, CERN-EP/99-129, accepted by Phys. Lett.

\bibitem{stat}
J.F. Grivaz and F. Le Diberder,
\newblock  Complementary analyses and acceptance optimization in new particle
  searches,
\newblock  Preprint Lal 92-37 (June 1992).

\bibitem{snow}
G. Anderson \etal,
\newblock  in DPF/DPB Summer Study on New Directions for High Energy Physics:
  Snowmass '96, Snowmass, CO, USA; 25 Jun - 12 Jul 1996,  (D.G. Cassel, L.
  Trindle-Gennari and R.H. Siemann, APS, New York, 1997), p. 669.

\bibitem{gstring}
C.H. Chen, M. Drees and J.F. Gunion,
\newblock  Phys. Rev. {\bf D 55}  (1997) 330,
\newblock  Erratum Phys. Rev. Lett. {\bf 82} (1999) 3192.

\bibitem{MSSM_GUT}
See for instance:\\ L. Ibanez, Phys. Lett. {\bf B 118} (1982) 73;\\ R.
  Barbieri, S. Farrara and C. Savoy, Phys. Lett. {\bf B 119} (1982) 343.

\bibitem{slepton}
L3 Collaboration, M. Acciarri \etal, CERN-EP/99-128, accepted by Phys. Lett.

\bibitem{caso}
C. Caso \etal, {\it Review of Particle Physics}, \EPJ {\bf C 3} (1998) 1.

\end{thebibliography}


\end{document}